\documentclass[twocolumn,showpacs,preprintnumbers,amsmath,amssymb,superscriptaddress]{revtex4}

\usepackage{graphicx}

\makeatletter
\def\@dotsep{5}
\makeatother

\begin{document}

\preprint{to be published on Appl. Phys. Lett. May, 29}

\title{A single intrinsic Josephson junction with double-sided fabrication technique}

\author{L. X. You}
\author{M. Torstensson}
\author{A. Yurgens}
\author{D. Winkler}

\affiliation{Quantum Device Physics Laboratory, Department of Microtechnology and Nanoscience (MC2), Chalmers University of Technology, SE-412 96 G\"oteborg, Sweden}

\author{C. T. Lin}
\author{B. Liang}
\affiliation{Max-Planck-Institut f\"ur Festk\"orperforschung, Heisenbergstrasse 1, D-70569 Stuttgart, Germany}

\date{\today}

\begin{abstract}
We make stacks of intrinsic Josephson junctions (IJJs) imbedded in the bulk of very thin ($d\leq 100$~nm) $\mathrm{Bi_2Sr_2CaCu_2O_{8+x}}$ single crystals. By precisely controlling the etching depth during the double-sided fabrication process, the stacks can be reproducibly tailor-made to be of any microscopic height ($0-9\ \mathrm{nm} <d$), i.e. enclosing a specified number of IJJ ($0-6$), including the important case of a single junction. We discuss reproducible gap-like features in the current-voltage characteristics of the samples at high bias.
\end{abstract}


\maketitle

Superconductor-insulator-superconductor (SIS) junctions are very important elements of nowadays superconducting electronics. Although fabrication techniques for low-temperature-superconductor SIS junctions are well developed, the high-temperature-superconductor (HTS) artificial SIS junctions have not yet reproducibly been made due to the short coherence length of HTS and problems of HTS thin-film technology.
The intrinsic Josephson effect that naturally exists in the layered HTS single crystals of very high anisotropy may provide a solution to this problem. The electric transport perpendicular to the layers in such materials occurs via the sequential tunneling of quasi-particles (and Cooper pairs) between the well conducting (and superconducting) layers across intermediate insulating ones. The effect was first experimentally revealed in 1992 by Kleiner \textit{et al} and Oya \textit{et al} in $\mathrm{Bi_2Sr_2CaCu_2O_{8+x}}$ (BSCCO)~\cite{Kleiner:PRL92, Oya:JJAP92}.  A BSCCO single crystal in the $c$-axis direction can thus be viewed as a large array of very thin (1.5~nm) intrinsic Josephson junctions (IJJs) in series.  At present, this is perhaps the only reliable way to obtain HTS SIS junctions. These junctions appear to have high potential for high-frequency applications like heterodyne receivers and quantum voltage standards~\cite{Wang:PRL01, Wang:APL02}.

It is difficult to isolate and study individual IJJ. Most of applications and research on IJJs use therefore stacks containing many junctions. The stacks can be formed either on the surfaces of single crystals (mesas) or etched out in the middle of BSCCO whiskers using focussed ion beams~\cite{Yurgens:review}. Earlier we developed a method for making a single intrinsic Josephson junction (SIJJ) enclosed in a U-shaped mesa structure~\cite{You:JJAP04}.  An effective SIJJ could be seen in the 4-probe measurements on such a mesa. However, extra junctions under the electrodes can give rise to unwanted local heating~\cite{You:SUST04, Wang:APL05}.

In this letter, we report on reproducible fabrication of uniform stacks formed  \textit{inside} the very thin ($d\sim 100$~nm) pieces of BSCCO single crystals by using a double-sided etching technique~\cite{Wang:APL01}. The number of junctions in the stacks can be tailor-made to any small number $N<10$ including the important case of $N=1$ (a single-junction stack). As the stacks are imbedded in the superconducting material, i.e. there are no direct resistive contacts to them, heating effects can be minimized in such stacks allowing the genuine current-voltage (I-V) characteristics to be traced out. Devices like SQUIDs are feasible to make using the developed technique.

The stack-fabrication routines start with evaporation of an Au thin film on a freshly cleaved crystal glued onto a sapphire substrate using polyimide. Using the conventional photolithography and Ar-ion etching, a bow-tie-shaped mesa with a micro-bridge in the center is formed on the crystal. The overall thickness of this mesa $d$ is typically about 100 nm which is controlled by etching time and rate. Then, a $40\ \mathrm{nm}\lesssim d/2$ deep slit is made across the bridge. In the next step, we flip the sample and glue it onto another sapphire substrate, sandwiching the single crystal between the two substrates. Separating the substrates cleaves the single crystal into two pieces, one with the mesa being flipped upside down and glued to the second substrate. We subsequently remove all material but the mesa by iteratively cleaving the former with the aid of Scotch tape and inspecting the resulting sample in optical microscope. A new gold layer is then deposited and patterned immediately after that to make four electrodes attached to this small piece of single crystal.  Finally, $\mathrm{CaF_2}$ protection layer with an open window placed across the bridge is formed using lift-off patterning. Further Ar-ion etching will incise a second slit into the bridge through that window.

\begin{figure}[width = 85 mu]
\includegraphics{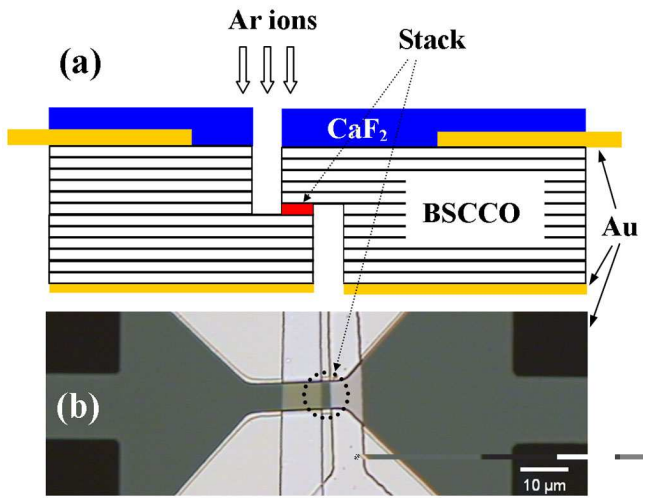}
\caption{(Color online) (a) Appearance of a single IJJ during consecutive etchings; (b) Optical images of a sample illuminated from the bottom; The red area between the two slits in (a) schematically shows the effective junction.}
\label{optics}
\end{figure}

From these processes there are two slits formed into the BSCCO bridge, one from the top and one from the bottom (see Fig.~\ref{optics}). The slits are separated by a certain distance along the bridge forming a zigzag structure. Initially, the sum-depth of the two slits is smaller than the overall thickness of the bridge. The upper slit can be made deeper with further etching and there will be a moment when the two slits overlap in the out-of-plane direction. At this moment, the current applied to the bridge will also flow across the newly formed stack of IJJs in the $c$-axis direction in the middle part between the slits.

By monitoring the resulting low-temperature ($T<T_c$) I-V's of the bridge after consecutive sessions of Ar-ion etching ($1-2$~minutes long, roughly corresponding to about $1-2$~nm of etched depth), we can predict and easily detect the moment when the slits overlap. This moment is characterized by a sudden decrease of the superconducting critical current and appearance of a distinctive hysteresis in the I-V's typical for the SIS-type IJJ. We take the sample out of the Ar-ion-beam chamber after each etching step to characterize the sample in the whole temperature range ($4.2-300$ K) \cite{note:etching}. In principle, the measurements could be done \textit{in-situ} by using a cryocooled sample holder in the chamber~\cite{Yurgens:APL}.

\begin{figure}
\includegraphics{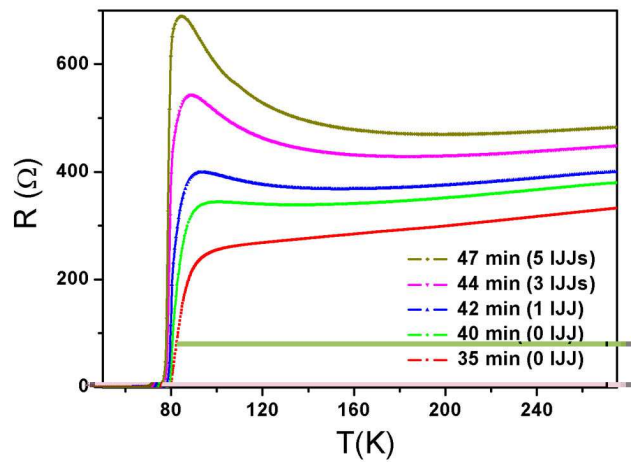}
\caption{(Color online) The $R(T)$-curves of the sample ($\mathrm{ 5.5\times 2.0\ \mu m^2}$) with increasing etching time. $T_c$ is almost the same for all the curves which demonstrates no evident degradation of the sample during the etching~\cite{note:etching}.}
\label{RT}
\end{figure}

\begin{figure}
\includegraphics{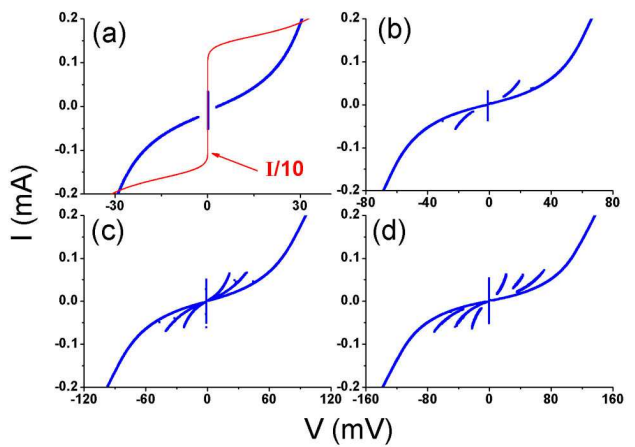}
\caption{(Color online) The I-V curves of the stack at 4.2~K with consecutively increasing number of junctions. The thin (red) curve in (a) shows the I-V curve before the appearance of the SIJJ. Note 10 times larger current scale for this curve.}
\label{IVs}
\end{figure}

Figs.~\ref{RT} and \ref{IVs} illustrate how the resulting resistance $R(T)$ and I-V's change as the second slit becomes deeper with the etching time. After some preliminary etching during 40 min (insufficient for the slits to overlap), the I-V is non-hysteretic and the critical current of the bridge is relatively high (see the thin curve in Fig.~\ref{IVs}(a)). In Fig.~\ref{RT} we also see a metallic $R(T)$-dependence (the second curve from bottom) which is characteristic for the $ab$-plane transport.

After two minutes of additional etching, the I-V drastically changes. The critical current $I_c$ drops from $\sim$ 1.3~mA to $\sim$ 50~$\mu$A and a hysteresis appears indicating that the slits overlapped and the SIJJ-stack formed (see the thick curve in Fig.~\ref{IVs}(a)). This $I_c$ corresponds to the $c$-axis critical current density of $j_c\sim 500$~A/cm$^2$ which is typical for our single crystals. In the rest of the panels of Fig.~\ref{IVs} we illustrate a progressive increase in number of I-V branches with the same $I_c$ corresponding to the gradual increase of the stack height. Moreover, the $R(T)$-curves change from metallic to semiconductor-like as the $c$-axis resistance of the stack becomes dominant in the overall resistance of the bridge. Several other samples showed very similar results.

It should be emphasized that despite the apparent simplicity, the technique provides us with the unprecedented $\sim$ 1~nm accuracy in controlling the stack height, thus assuring any low number of IJJs in the stack. It is also the first time when a SIJJ is controllably and reproducibly made using the double-sided fabrication technique.

\begin{figure}
\includegraphics{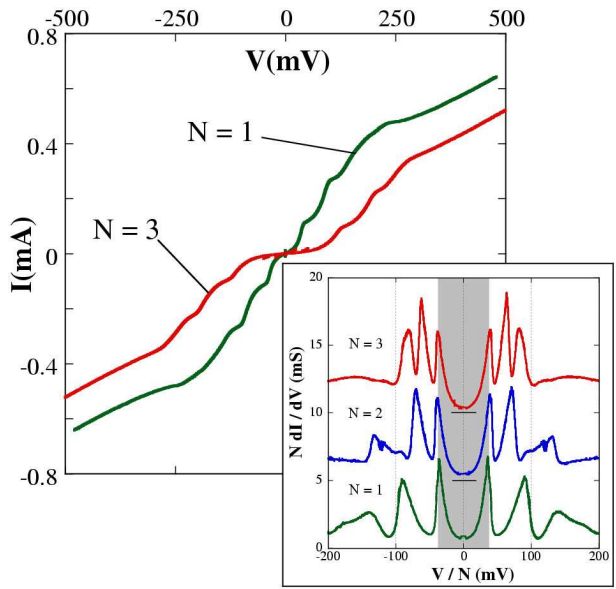}
\caption{(Color online) $I(V)$- and $dI/dV(V/N)$- curves of another sample ($1.7\times 1.7\ \mathrm{\mu m}^2$) at 4.2~K. The gray rectangular marks the voltage where scaling of the assumed gap is observed.}
\label{SIJJ}
\end{figure}

Several features can be seen in I-V's while increasing the bias current (see Fig.~\ref{SIJJ}).  $dI/dV(V)$-curves therefore contain several peaks that behave differently with the stack height ($N$). High-bias peaks usually shift in normalized voltage ($V/N$), i.e. their positions depend on $N$ (no ``scaling"; see the inset to Fig.~\ref{SIJJ}). The low-lying peaks at $V/N\approx \pm 36$~mV are however $N$-independent.  These peaks are reproducibly found in all five stacks fabricated so far, of various lateral dimensions ($\mathrm{ 0.5-11\ \mu m^2}$), with slightly different critical temperatures ($T_c=60-80$~K), and with different $N$. Averaged over 25 measurements on all the stacks, the voltage of the peaks is $V_p\approx 32.4$~meV at 4.2~K with standard deviation of 4~meV. No clear correlation between $V_p$ and $T_c$ could firmly be established, except for the sample with the smallest $V_p=28$~meV that had the lowest $T_c=60~K$.   These peaks persist in a wide temperature range ($T\leqslant 0.8T_c$) and for all low $N\leqslant 5$, although becoming weaker with $T$ and $N$.

The recurrence of the observed feature suggests that it might correspond to the superconducting order parameter $\Delta = V_p/2 \approx 14-18$~meV. Although being significantly smaller than in the majority of STM data ($30-50$~meV)~\cite{Renner:PRB95,Kapitulnik:PRL05,disorder}, the suggested $\Delta \approx 14-18$~meV is not totally odd. In fact, similar values were repeatedly measured in break-junction experiments on overdoped samples~\cite{Zasadzinski}, as well as in point-contact measurements of the gap anisotropy~\cite{Kane:anisotropy}.

In spite of the reproducibility of the results, we cannot exclude the possibility that the peaks are due to, e.g. the current-driven transition of a few surface CuO planes in the thinned regions of the bridge~\cite{You:PRB05}. Such a transition would give rise to breaks in I-V's and correspondingly - to peaks in $dI/dV(V)$. Indeed, taking the sheet critical current density of $0.3-0.7$~A/cm~\cite{You:PRB05} and the width of the bridge (1.7~$\mu$m), we get a range of critical currents ($0.05-0.12$~mA) which obviously includes the current at which the first break is seen in Fig.~\ref{SIJJ}. A more complex geometry of samples with four superconducting arms reaching to the stack in the middle is currently designed to avoid the possible contributions from such transitions.

The in-plane gap anisotropy~\cite{Kane:anisotropy} and Joule heating might be other possible explanations for the small gap. However, it is not clear how the supposedly d-wave anisotropy of the gap could result in just one half of the maximum value for the out-of-plane transport. The Joule heating can easily be ruled out as the power dissipation is negligibly small. It is only $4-12\ \mu$W at the peak positions (see Fig.~\ref{SIJJ}) resulting in $<0.5$~K of overheating assuming the overall thermal resistance of $\sim$50~K/mW~\cite{Yurgens:PRL04}.

In summary, we successfully combine the double-sided fabrication method with precise control over etching to make stacks with any low number of intrinsic Josephson junctions, including the single intrinsic Josephson junction. In addition to intrinsic tunneling spectroscopy of superconducting gap, the technique is also important for possible applications like SQUDs and high-frequency mixers.

This work is financed by the Swedish Foundation for Strategic Research (SSF-OXIDE), and the Swedish Research Council (grants 6212002-4995 and 6212002-5180).


\newpage



\begin{thebibliography}{99}
\bibitem{note:etching} Ar-ion etching was found to suppress the superconductivity in the slit region. It could be easily recovered by warming the sample up to 100$^\circ$C in air.
\end{thebibliography}
\end{document}